\documentclass[article,
 amsmath,amssymb,
 aps,
]{revtex4-2}

\usepackage{graphicx}
\usepackage{bm}

\usepackage{epstopdf} 

\begin{document}

\title{Crossing of phantom divide line in model \\of interacting Tsallis holographic dark energy}

\author{Artyom V. Astashenok}
 \email{aastashenok@kantiana.ru.com}
\author{Alexander S. Tepliakov}
\affiliation{%
I. Kant Baltic Federal University\\
236041, Kaliningrad, Russia, Nevskogo str. 14
}%

\begin{abstract}
We consider a Tsallis holographic dark energy model with interaction between dark energy and matter. The density of dark energy is taken as $\rho_d \sim 3C^2/L^{4-2\gamma}$, where $C$, $\gamma$ are constants. The event horizon is chosen as the characteristic scale $L$. The cosmological dynamics of the universe are analyzed, with special attention paid to the possibility of crossing the phantom line $w_{eff}=-1$. It is shown that for certain values of parameters this may occur not only once, but also twice.
\end{abstract}
\maketitle
\section{Introduction}
Our universe is expanding with acceleration ~\cite{1,2}. Type Ia supernovae in distant galaxies, the distance to which has been determined by Hubble's law, have a brightness lower than that obtained by the expansion rate of the universe filled with matter and radiation.
The universe began to expand with acceleration about 5 billion years ago. The reason for this acceleration is most likely the so-called ``dark energy''. It is known that dark energy is distributed in space with an extremely high degree of homogeneity, has a low density and does not interact appreciably with ordinary matter through the known fundamental types of interaction---with the exception of gravity. There are many dark energy models, but the best agreement with observational data on Type Ia supernovae~\cite{Amanullah}, baryonic acoustic oscillations~\cite{Blake} and the Hubble redshift dependence of the Hubble constant is seen for the simple $\Lambda$CDM model~\cite{LCDM-1, LCDM-2, LCDM-3, LCDM-4, LCDM-5, LCDM-6, LCDM-7}. In the $\Lambda CDM$ model, the nonzero cosmological constant plays the role of dark energy. However, from the point of view of quantum field theory, the cosmological constant should be 120 orders of magnitude higher than its observed value. This is the mystery of the smallness of the cosmological constant. 

There are two other options for explaining the nature of dark energy. According to the first, the dark energy is the so-called quintessence---a scalar field with the effective parameter of the state, the value of which lies in the range $-1<w<-1/3$~\cite{Caldwell, Steinhardt, Ferramacho, Caldwell-2}. Unlike the cosmological constant, the quintessence is a dynamic field, and its energy density depends on time.

In the models of modified gravity~\cite{Capozziello, Odintsov, Turner}, it is assumed that the accelerated expansion of the universe is caused by deviation of the gravity model from GR on cosmological scales. 

In recent years, the holographic dark energy model has been actively investigated (see the review~\cite{Wang} and references therein). Its theoretical basis is the holographic principle (see~\cite{3, 4, 5}) and its various modifications. The holographic principle states that all physical quantities within the universe, including the dark energy density, can be described by setting some quantities at the boundary of the universe. This leaves only two physical quantities through which the dark energy density can be expressed: the Planck mass $M_{p}$ and the characteristic scale $L$. For $L$, one can take the particle horizon, the event horizon or the inverse of the Hubble parameter. 

Tsallis and others  generalized the Boltzmann--Gibbs entropy for a black hole to the nonextensive entropy~\cite{Tsallis, Lyra, Wilk}, and the Tsallis holographic dark energy (THDE)~\cite{Tsallis-2} model was proposed. In this model, the horizon entropy of a black hole is $S\sim a A^{\delta}$, where $a$ is an unknown constant and $\delta$ is the nonadditivity parameter. The classical Bekenstein result for entropy takes place for $\delta=1$. The energy density for the THDE density can usually be written as $\rho\sim H^{4-2\delta}$, where the Hubble parameter plays the role of the infrared cutoff. 

Applications of this model are considered in many papers. In the context of cosmological models, THDE is investigated in a flat universe without interaction between matter and dark energy~\cite{Bamba}. In \cite{Saridakis}, the authors considered nonextensive thermodynamics with a varying exponent. In the framework of modified gravity, the Tsallis model has been discussed in terms of Brans--Dicke gravity~\cite{Lobo, Jawad}, dynamical  Chern--Simons gravity~\cite{Rani}, $f(T)$ and $f(G; T)$ gravity~\cite{Aly, Sharif} and many others. We considered this model and its generalization to cosmology on the Randall--Sundrum brane in the paper~\cite{AA}, where we investigated the admissible values of the model parameters at which it agrees with the astrophysical observational data. 

Note that THDE is a particular example of generalized Nojiri--Odintsov HDE introduced in~\cite{Nojiri:2005pu, Nojiri:2017opc}. This was explicitly demonstrated in~\cite{Nojiri:2021iko, Nojiri:2021jxf}.  

Arguments in favor of THDE model can also be found regarding quantum gravity of black holes~\cite{Barrow}. The consequences and implications of this generalized entropy in cosmological setups have been studied. This shows that the generalized entropy may be in accordance with the thermodynamics laws, the Friedmann equation and universe expansion. 

In this paper, we will consider the THDE model with the inclusion of a possible interaction between matter and dark energy. We will analyze a simple type of interaction ($\sim$$H\rho_{de}$). The interaction leads to the decay of dark energy and at a certain intensity the fractions of matter and dark energy in the overall balance stabilize. As a result, we have a quasi-de Sitter expansion of the universe instead of a possible singularity in a future. 

The paper is organized in the following way. In Section II, a model of Tsallis holographic dark energy with interaction is described. Then, we analyze a possible phantom divide line crossing for dark energy. We mainly consider the case when the scale cutoff for THDE is the event horizon. Other possibilities such as particle horizon and inverse Hubble parameter are also briefly investigated. In the last section, some concluding remarks are presented.

\section{Model Description}

Consider a spatially flat universe with the Friedman--Lemaitre--Robertson--Walker metric:
\begin{equation} 
\label{eq:1}
ds^2=dt^2-a^2(t)(dx^2 + dy^2 + dz^2).
\end{equation}

Here, $t$ is cosmic time, $a(t)$ is the scale factor. Let us assume that the universe is filled with dark energy and matter with densities $\rho_{de}$ and $\rho_m$, respectively. The cosmological equations for a given metric can be written in the following form: 
\begin{equation}
\label{eq:2}
H^{2} = \frac{1}{3}(\rho_{m}+\rho_{de})
\end{equation}
\begin{equation}
\label{eq:3}
\dot{H}+H^{2} = -\frac{1}{6}(\rho_{m}+\rho_{de}+p_{de}).
\end{equation}

In the THDE model, the dark energy density is
\begin{equation}
\label{eq:4}
\rho_{de}=\frac{3C^{2}}{L^{4-2\gamma}},
\end{equation}
where $\gamma \in [1,2]$. The value $\gamma = 1$ corresponds to a simple holographic dark energy model, at $\gamma=2$ we have the cosmological constant. As a scale $L$, we can consider the event horizon:
\vspace{12pt}$$L = a \int_{t}^{\infty}\frac{d{t}'}{a}.$$

Other possibilities are the particle horizon
$$
L = a \int_{0}^{t}\frac{d{t}'}{a}
$$
and the inverse Hubble parameter
$$
L = 1/H.
$$

If dark energy and matter interact with each other, the continuity equations for the corresponding components take the form:
\begin{equation}
\label{eq:5}
\dot{\rho}_{m}+3H\rho_{m}=Q,
\end{equation}
\begin{equation}
\label{eq:6}
\dot{\rho}_{de}+3H(\rho_{de}+p_{de})=-Q.
\end{equation}

\begin{figure}[h]
\includegraphics[scale=0.25]{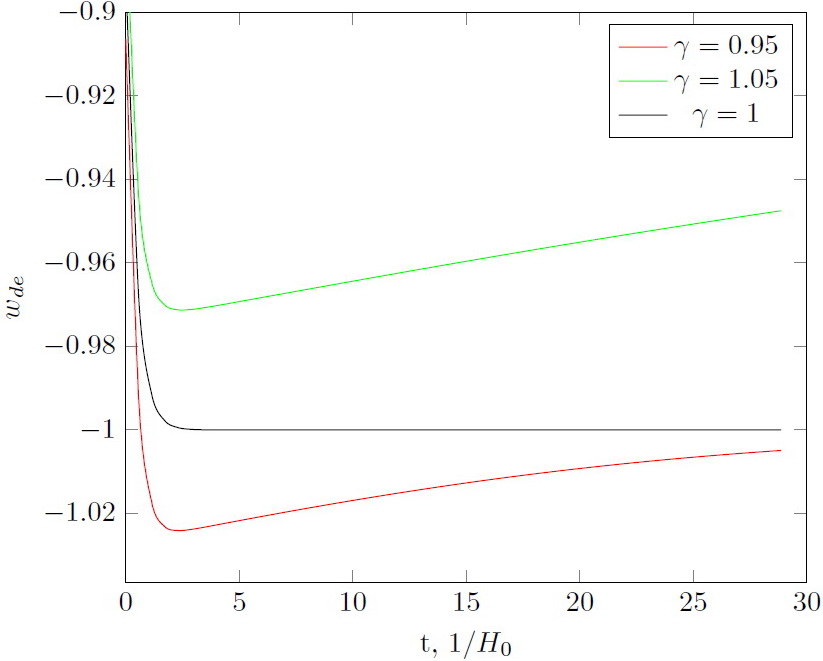}\includegraphics[scale=0.25]{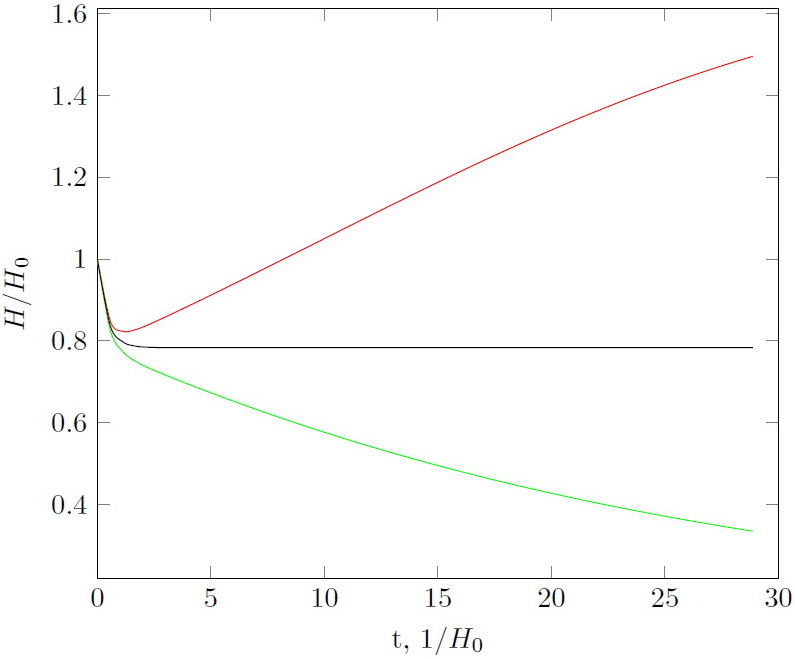}\\
\includegraphics[scale=0.25]{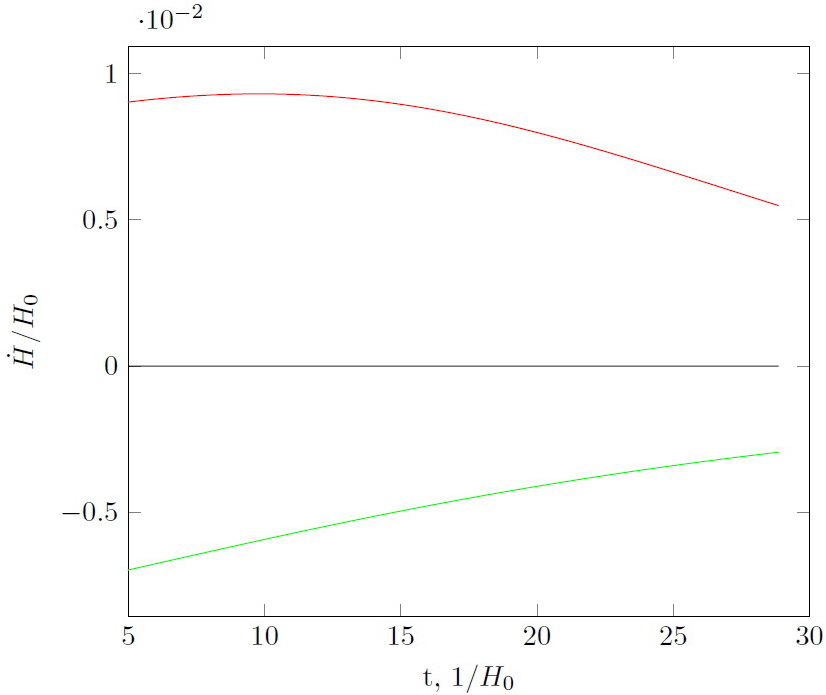}\includegraphics[scale=0.25]{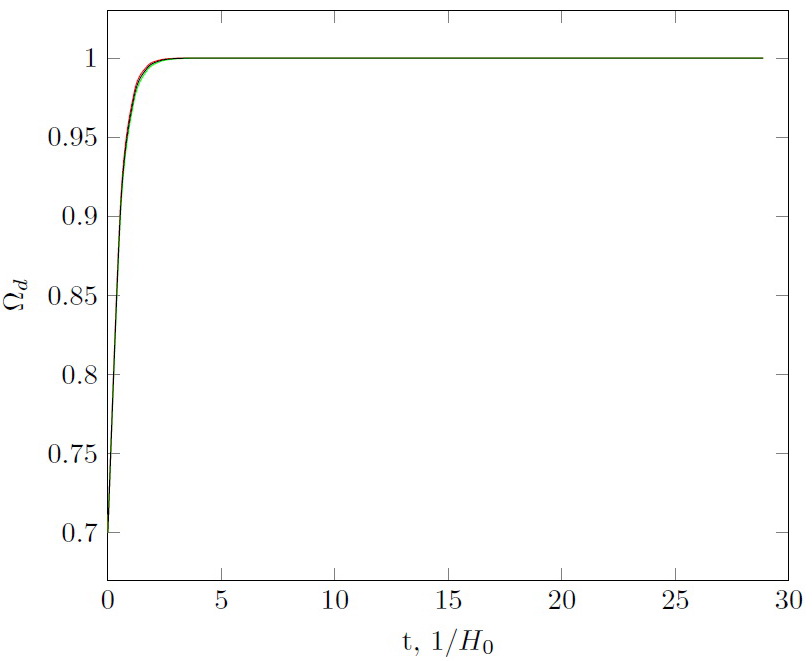}
\caption{The effective state parameter for holographic dark energy (\textbf{left, top}), the Hubble parameter (\textbf{right, top}) and the derivative of the Hubble parameter (\textbf{left, bottom}) and part of dark energy in total energy density (\textbf{right, bottom}) as a function of time for $C=1$, $d=0$. Time is given in units of $1/H_{0}$, where $H_0$ is the value of the Hubble parameter at the present time.}
    \label{fig11}
\end{figure}

Here, some function $Q$, which generally depends on time and densities, is introduced into the right-hand sides of the equations. The total density $\rho=\rho_{de}+\rho_m$ satisfies the usual continuity equation. Numerically solving the equations above, we can express the density of matter $\rho_{m}$ and analytically obtain an expression for the pressure of dark energy $p_{de}$. 
\begin{equation}
\label{eq:7}
p_{de}=-\frac{\dot{\rho}_{de}}{3H}-\rho_{de}-\frac{Q}{3H}.
\end{equation}

Then, the parameter of the dark energy equation of state is 
\begin{equation}
\label{eq:8}
w_{de}=\frac{p_{de}}{\rho_{de}}=-\frac{\dot{\rho}_{de}}{3H\rho_{de}}-1-\frac{Q}{3H\rho_{de}}.
\end{equation}

We will also analyze the evolution of $\Omega_m$ and $\Omega_{de}$, the fractions of matter density and dark energy in the future: 
\begin{equation}
\label{eq:9}
\Omega_{i}=\frac{\rho_{i}}{3H^{2}}.
\end{equation}

In the case of event or particle horizon as scale cutoff, we also add the following equation for $L$:
\begin{equation}
    \frac{d}{dt}\frac{L}{a}=\pm 1/a,
\end{equation}
where `$\pm$' signs correspond to particle and event horizon, correspondingly.

\section{The Possibility of Phantom Line Crossing and Disappearing of Singularities Due~to~Interaction}

In this section, we consider the case of an event horizon for $L$ and a simple choice for $Q$ in the form 
\begin{equation}
Q=3d^2 H\rho_{de},\quad {d}=\mbox{const}.
\end{equation}

Here and below, we assume $\Omega_{de}=0.7$ as the initial value for the fraction of dark energy. Numerical calculations show that for some values of $\gamma$, $C^2$ and the coupling constant $d^2$, the line $w_{de}=-1$ can be crossed for holographic dark energy. 

For $C=1$, $\gamma=1$ and in the absence of interaction between matter and dark energy, it is known that the expansion of the universe over long periods occurs according to de Sitter's law (see Figure \ref{fig11}). The same scenario takes place for $\gamma<1$, but after some transition period phantomization occurs, the value of state parameter goes to minimum and then increases and tends to $-1$ from below. For $\gamma>1$, the Hubble parameter approaches $0$ for $t\rightarrow\infty$ and expansion asymptotically stops.   

\begin{figure}[h]
\centering 
    \includegraphics[scale=0.25]{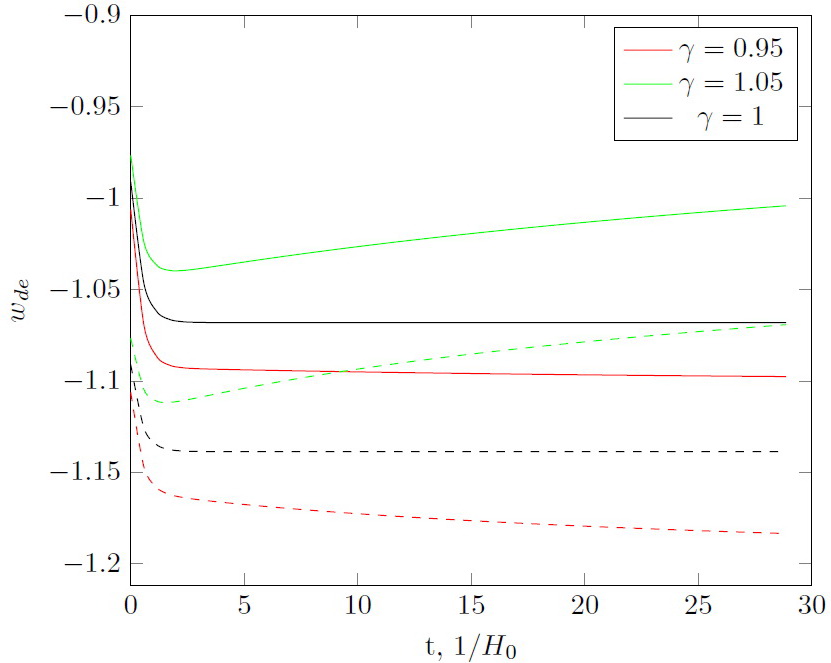}\includegraphics[scale=0.25]{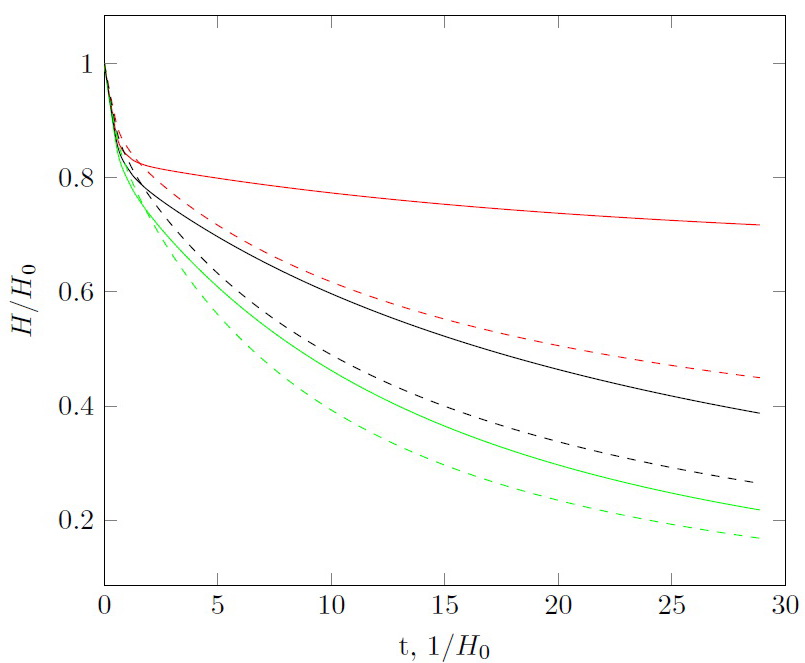}\\
    \includegraphics[scale=0.25]{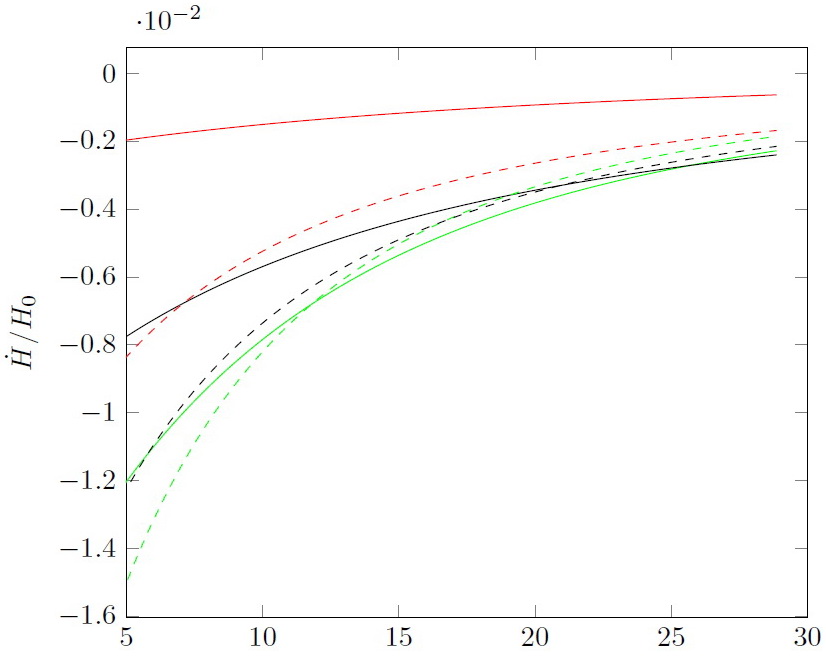}\includegraphics[scale=0.25]{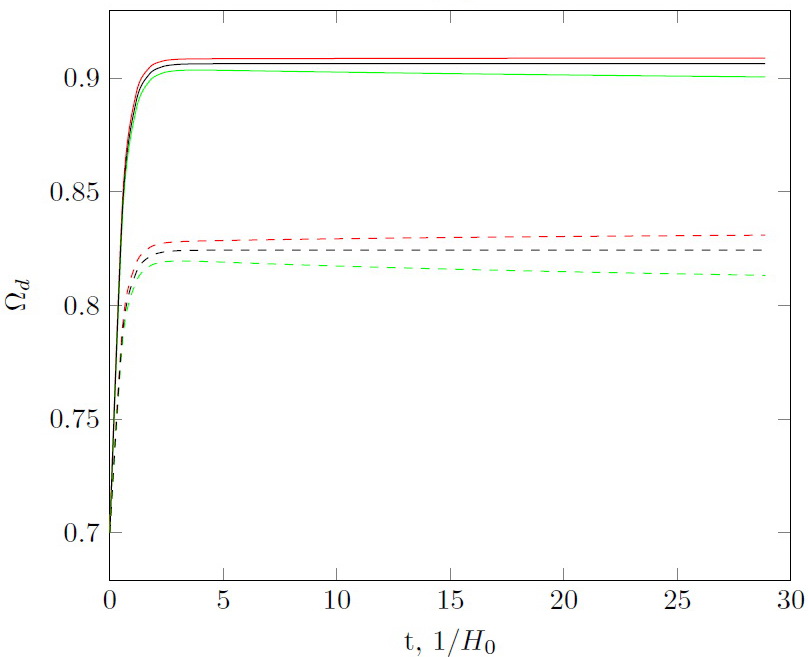}
    \caption{Same as in Figure~\ref{fig11}, but for $C=1$, $d^2=0.1$ (solid lines) and $d^2=0.2$ (dashed lines). For $\gamma=1.05$ and $d^2=0.1$, the value of $w_{de}$ intersects the phantom divide line $w_{0}=-1$ twice. Asymptotical values of $w_{de}$ for $\gamma=1$ and $\gamma=0.95$ are less than $-1$ but we have no singularities in the future and the universe expands according to the de Sitter law at $t\rightarrow\infty$.}
    \label{fig12}
\end{figure}

For the same value of $C$ and $\gamma>1$, but at $d^2>0$, the value of the state parameter $w_{de}$ can intersect the phantom line $w_{ph}=-1$ (see Figure \ref{fig12}). The value of the minimum depends upon $d^2$. After the minimum, the value of $w_{de}$ increases and the Hubble parameter tends to zero $t\rightarrow\infty$. For some $d^2$, the phantom divide line $w_{0}=-1$ may be crossed again. $H\rightarrow 0$ for $t\rightarrow 0$, therefore we have phantom energy without accelerated expansion. 

The cosmological evolution for $\gamma=1$ and $\gamma<1$ is similar. The parameter of state $w_{de}$ tends to $w_{0}<-1$ (for $\gamma=1$ this asymptotic is reached faster) but the future evolution does not contain the singularities one would expect for phantom energy. The Hubble parameter tends to a constant value, i.e., the expansion of the universe proceeds in a quasi-de Sitter regime.

The increase in $d^2$ leads to decreasing of the minimal value of $w_{de}$ and asymptotical values of $w_{0}$ for $\gamma\leq 1$ (see Figure~\ref{fig12}).

\begin{figure}[h]
\centering 
    \includegraphics[scale=0.25]{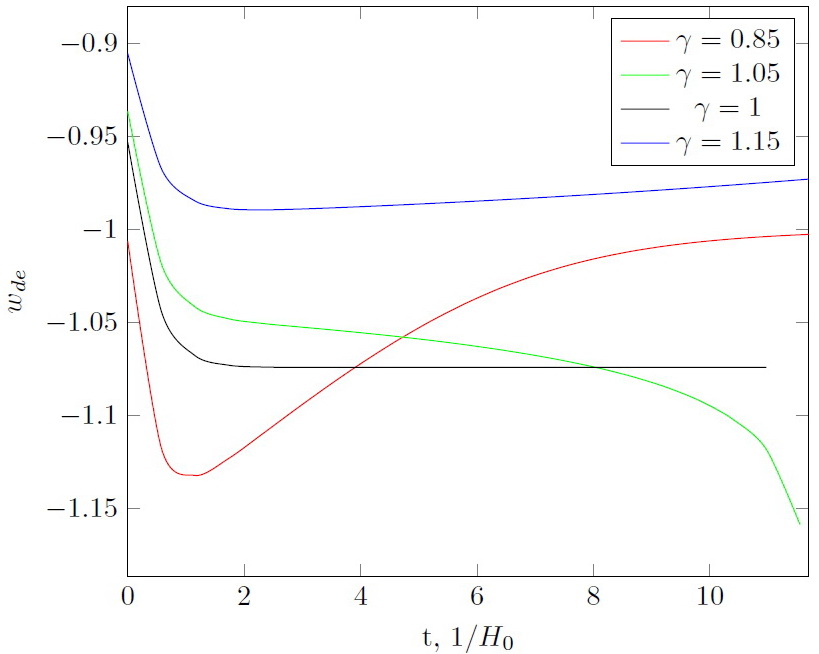}\includegraphics[scale=0.25]{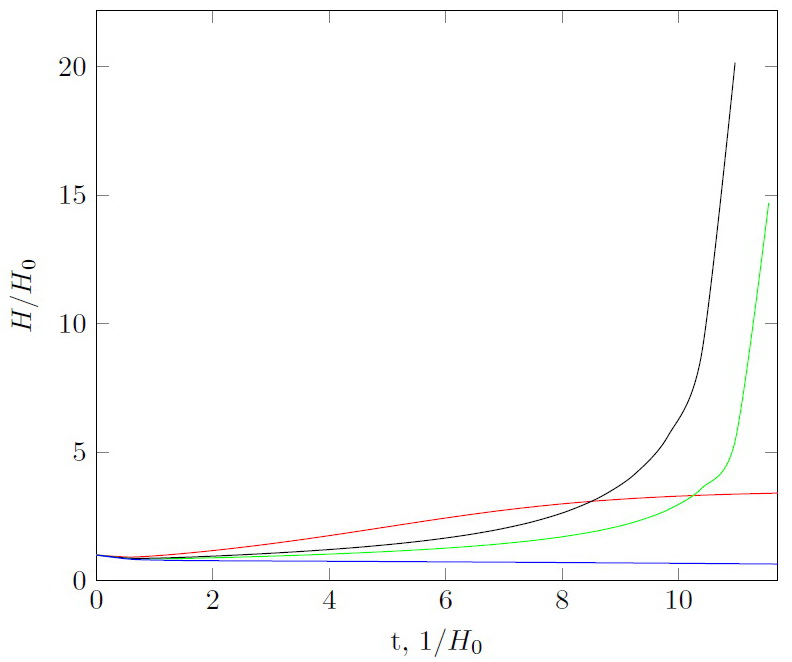}\\
    \includegraphics[scale=0.25]{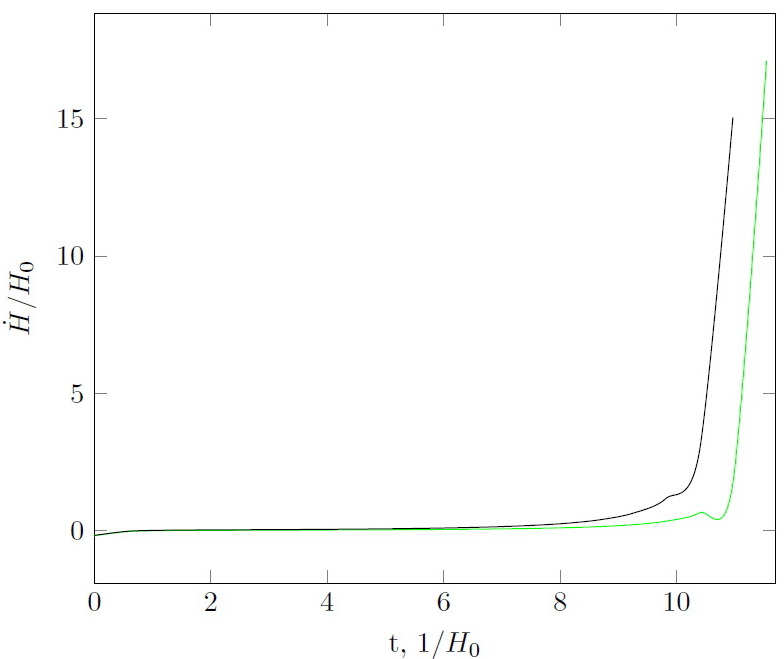}\includegraphics[scale=0.25]{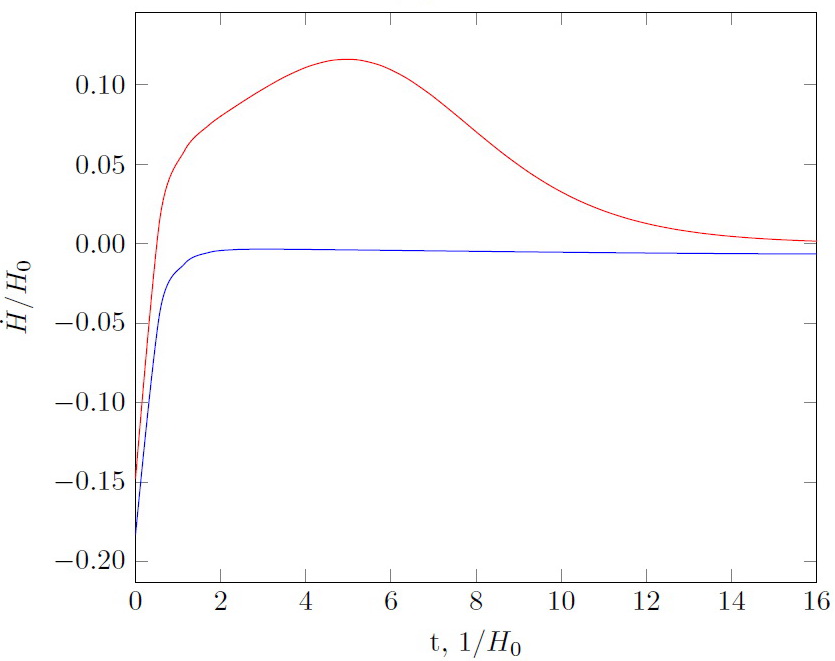}\\
    \includegraphics[scale=0.25]{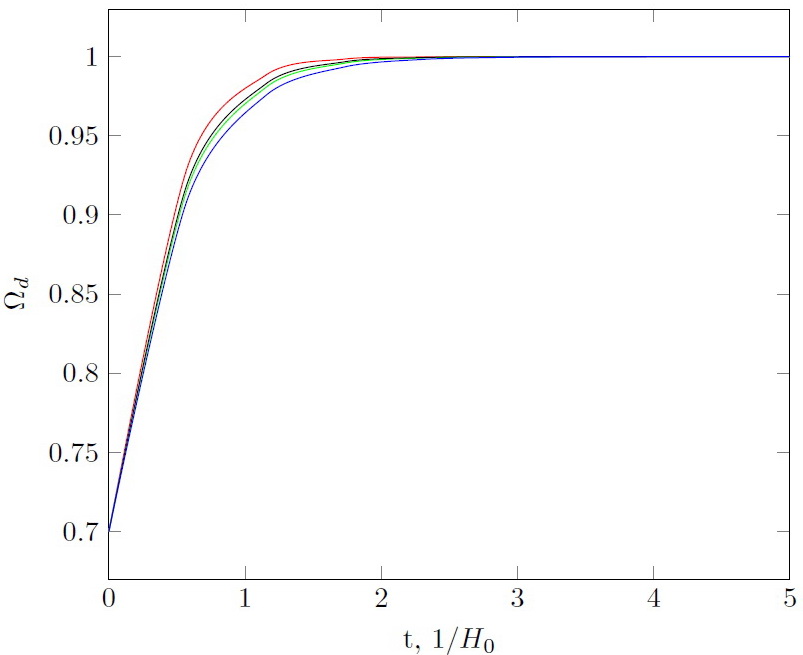}
    \caption{Same as in Figure~\ref{fig11}, but for $C=0.9$, $d=0$. For some $\gamma\geq 1$, big rip singularity in future takes place (green curve). For another $\gamma>1$, $H\rightarrow 0$. For $\gamma<1$, we have quasi-de Sitter expansion ($w_{de}\rightarrow -1$) in future.}
    \label{fig14}
\end{figure}

Next, consider the value of $C<1$. For $\gamma=1$, the cosmological evolution of the universe is the same as in the case of the phantom field with a constant equation-of-state parameter and therefore a dark energy big rip singularity occurs in the future (Figure~\ref{fig14}). For $\gamma<1$, the universe asymptotically expands according to the de Sitter law.  Two variants are possible for $\gamma>1$. If $1<\gamma<\gamma_0$ where $\gamma_0$ is some limit for given $C$, then the value of $w_{de}$ crosses $-1$ and tends to $-\infty$. The universe ends its existence in a big rip singularity. For $\gamma>\gamma_0$, we again have a quasi-de Sitter expansion in the future.

\begin{figure}[h]
\centering 
    \includegraphics[scale=0.25]{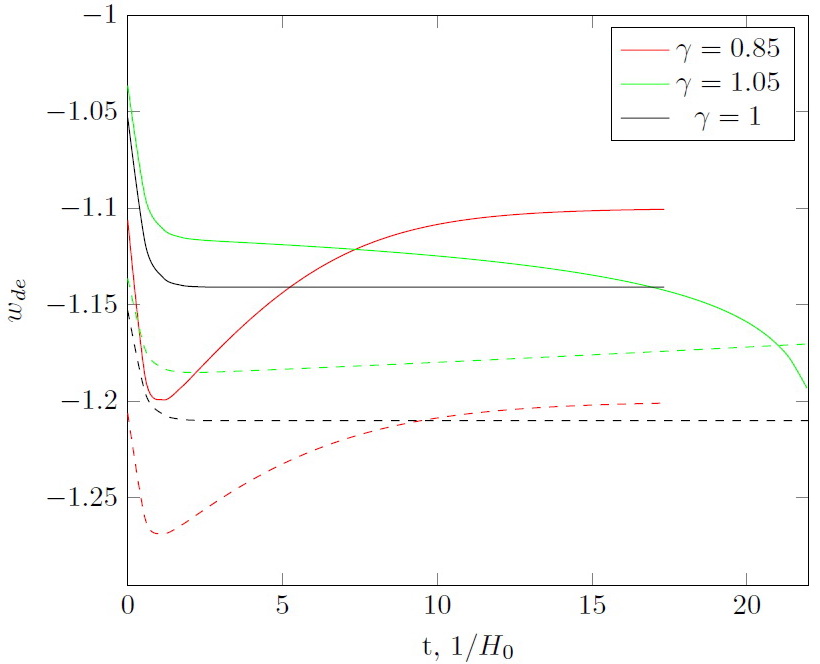}\includegraphics[scale=0.25]{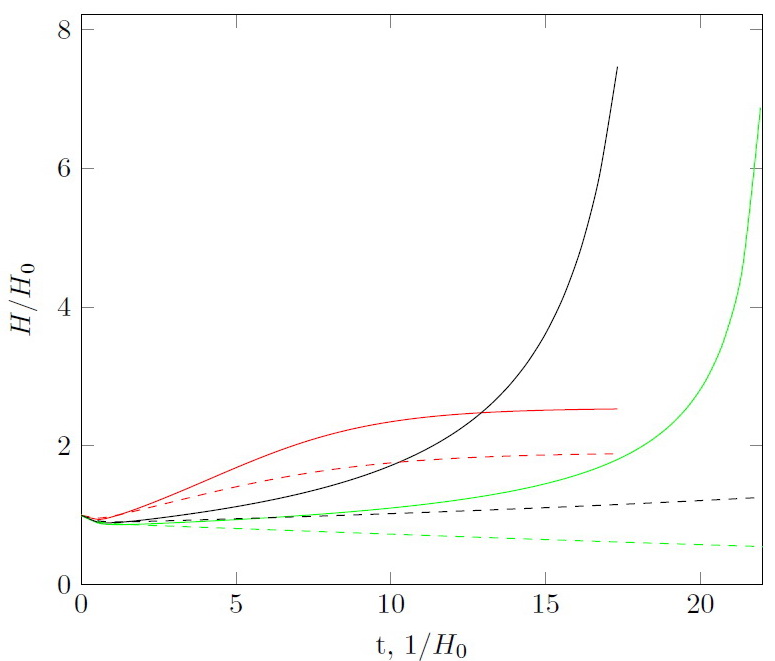}\\
    \includegraphics[scale=0.25]{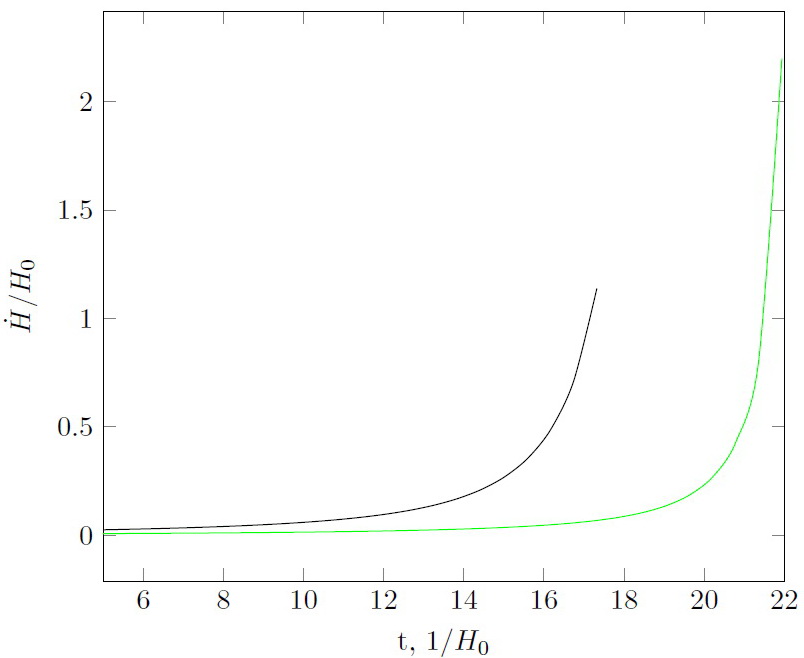}\includegraphics[scale=0.25]{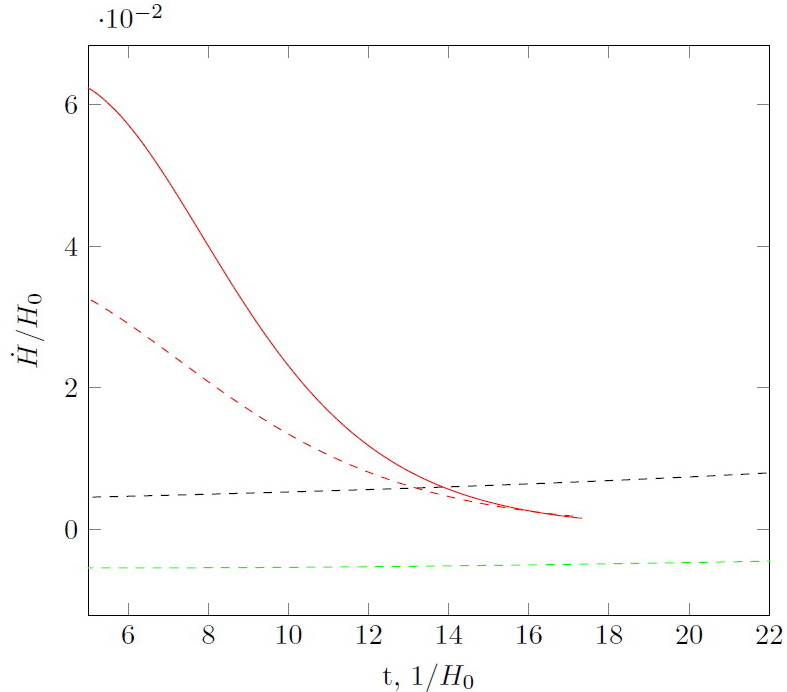}\\
    \includegraphics[scale=0.25]{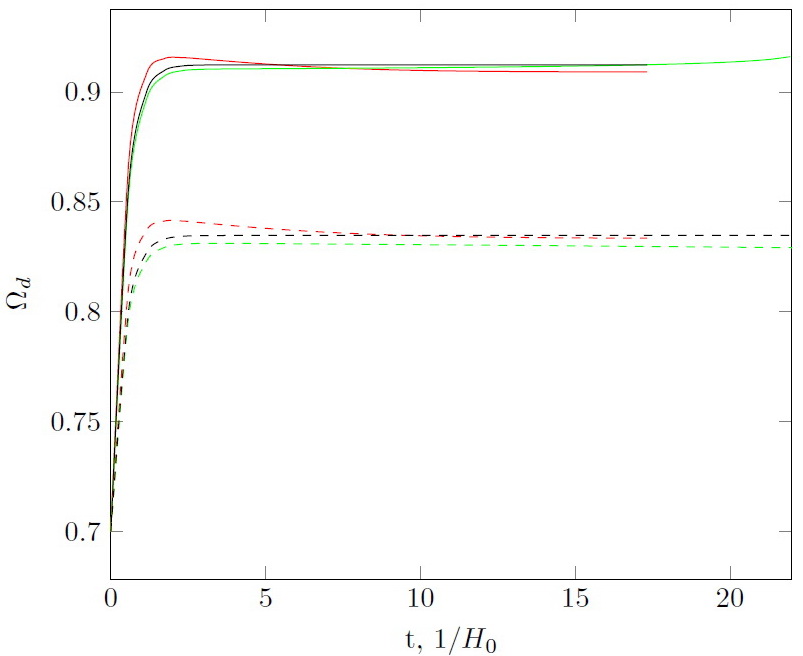}
    \caption{Same as in Figure~\ref{fig11}, but for $C=0.9$, $d^2=0.1$ (solid lines) and $d^2=0.2$ (dashed lines). For $\gamma=1.05$ at $d^2=0.2$, there is no big rip singularity which takes place for $d^2=0.1$ and without interaction.}
    \label{fig15}
\end{figure}

The interaction leads to the appearance of features of the same kind as for $C=1$ (Figure~\ref{fig15}). The asymptotic value of the state parameter at $\gamma=1$ is greater in comparison to the case without interaction. However, a big rip singularity occurs later. For $\gamma<1$, the interaction leads to asymptotical de Sitter expansion with $\dot{H}\rightarrow 0$ {over long periods} 
but with $w_{de}<-1$. A similar situation occurs for $\gamma>1$ although $w_{de}\rightarrow\-\infty$. 

\begin{figure}[h]
\centering 
    \includegraphics[scale=0.25]{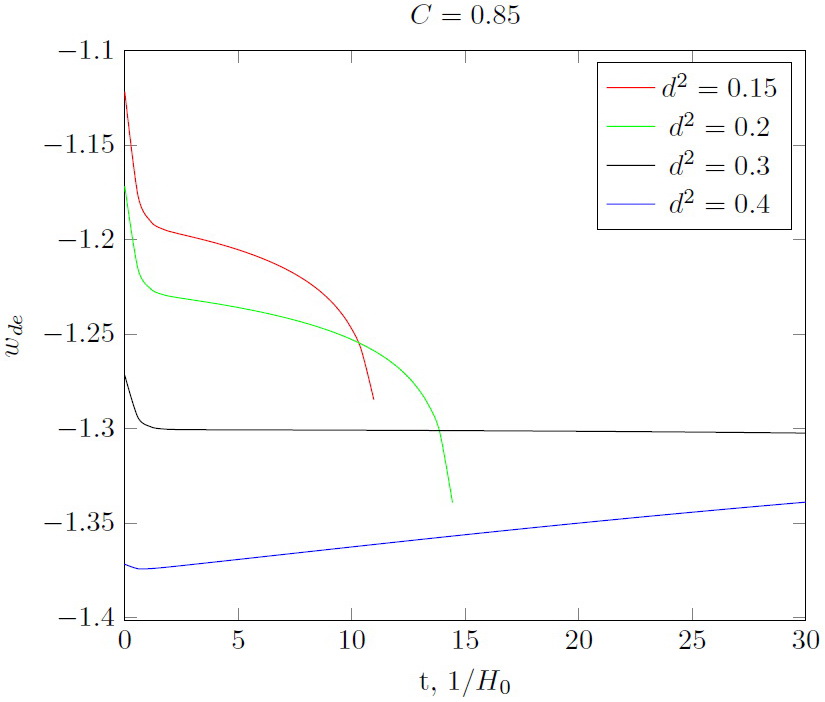}\includegraphics[scale=0.25]{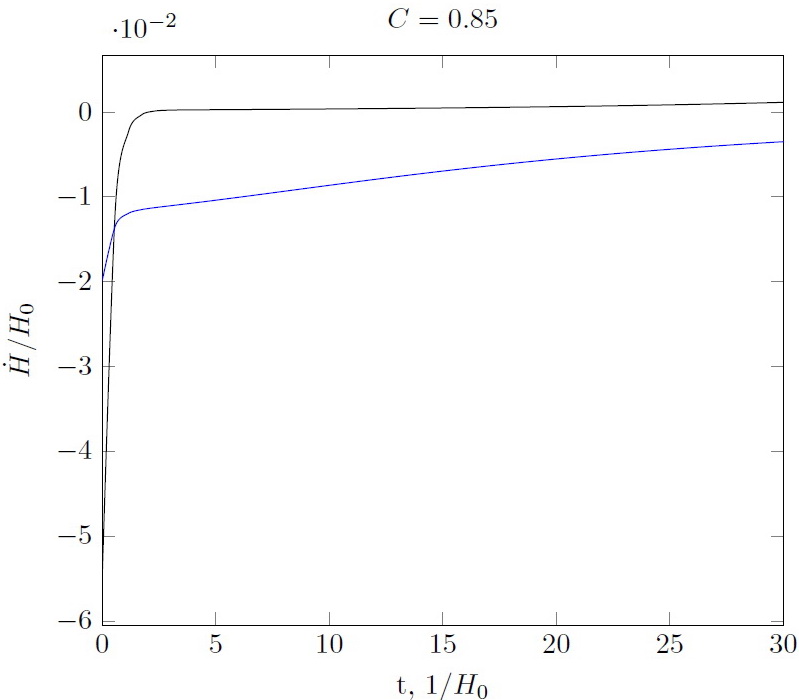}\\
    \includegraphics[scale=0.25]{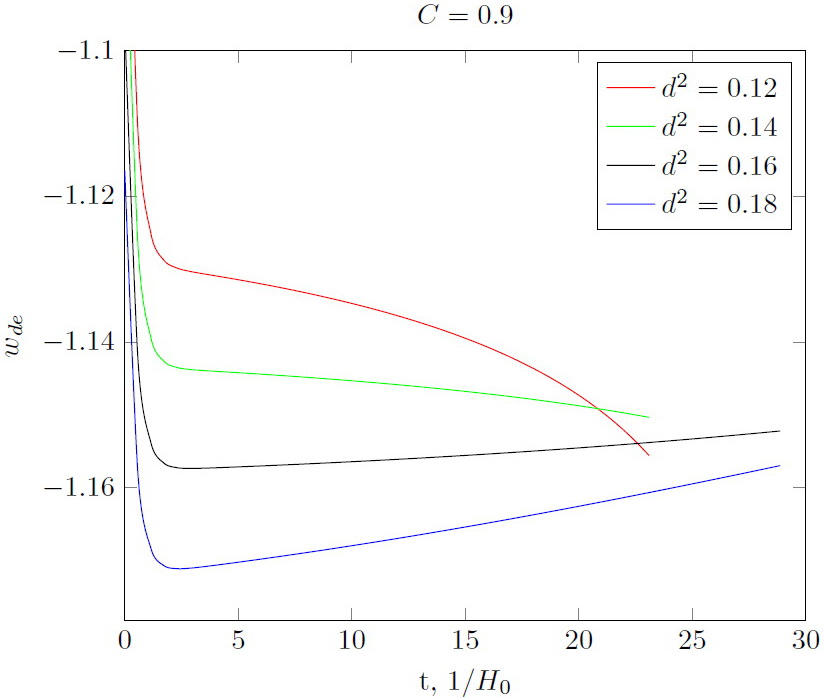}\includegraphics[scale=0.25]{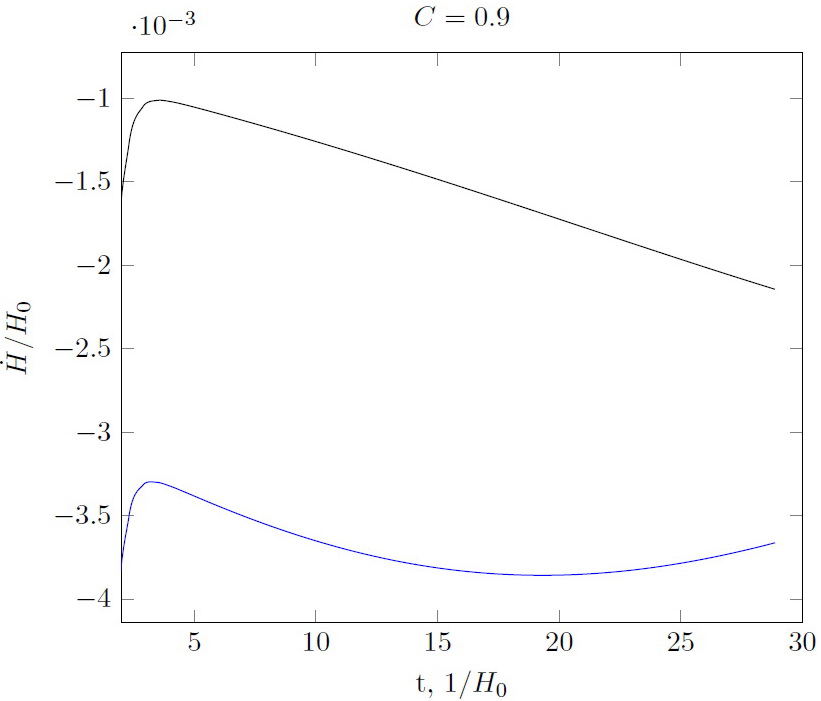}
    \caption{Evolution of the equation-of-state parameter and first derivative of the Hubble parameter for two values of $C^2$ in the case of $\gamma=1.05$ for various $d^2$. The value of the state parameter either ``skips'' into the negative region (slower than at $d^2=0$), or reaches the minimum $w_{min}<-1$ and starts growing (then there is no singularity in the future, the Hubble parameter decreases with time). These two cases are again separated by the value of $d^2_{crit}$ for specific $C$ and $\gamma$. }
    \label{fig16-2}
\end{figure}

One can see that an increase in $d^2$ eliminates the big rip singularity for $\gamma>1$ (Figure~\ref{fig15}) but not for $\gamma=1$. Further analysis shows that there is a critical value of $d^2$, above which the big rip singularity does not occur for $\gamma>1$ and the derivative of the Hubble parameter tends to 0 for  $t\rightarrow\infty$ (Figure~\ref{fig16-2}). For $d^2<d^2_{crit}$, the value of the Hubble parameter initially decreases and then begins to increase.

\begin{figure}[h]
\centering 
 \includegraphics[scale=0.25]{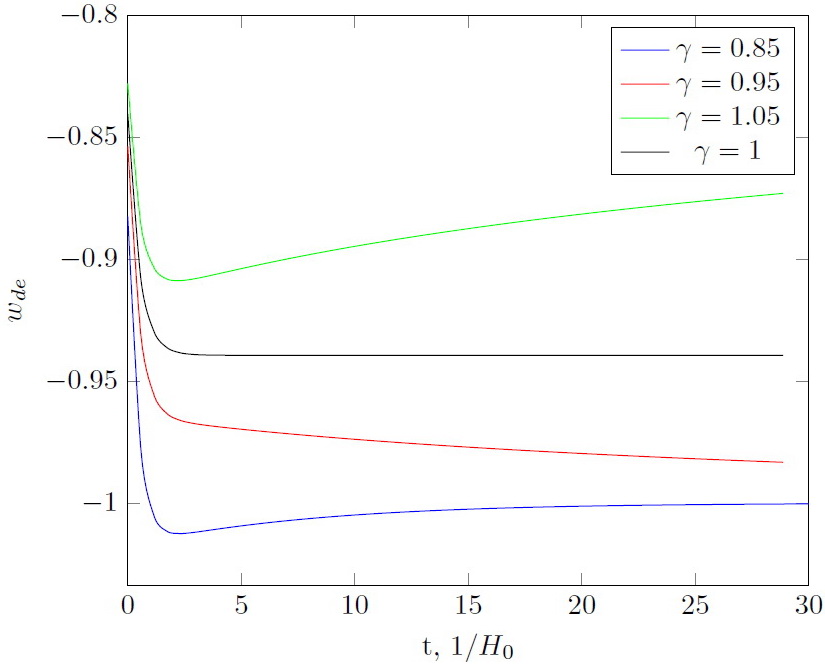}\includegraphics[scale=0.25]{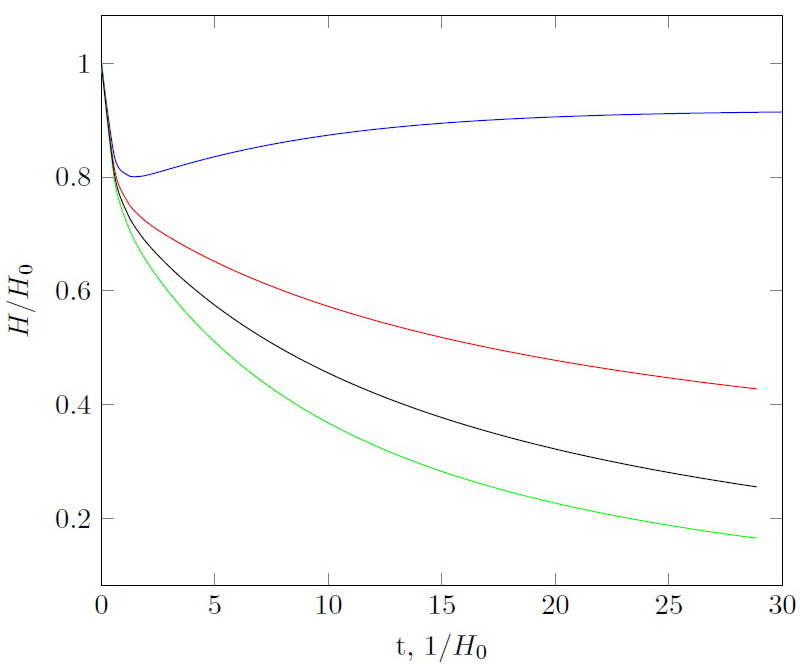}\\
    \includegraphics[scale=0.25]{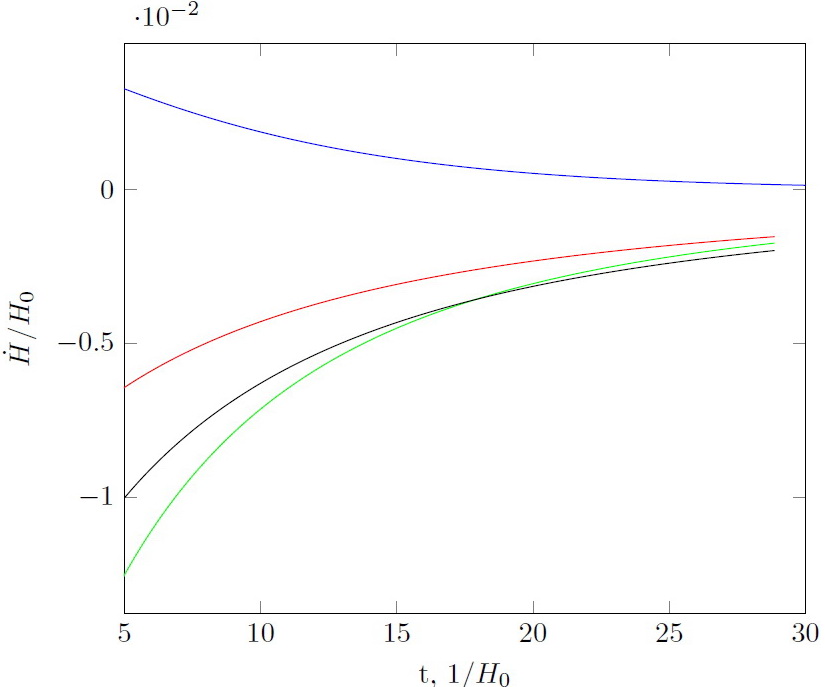}\includegraphics[scale=0.25]{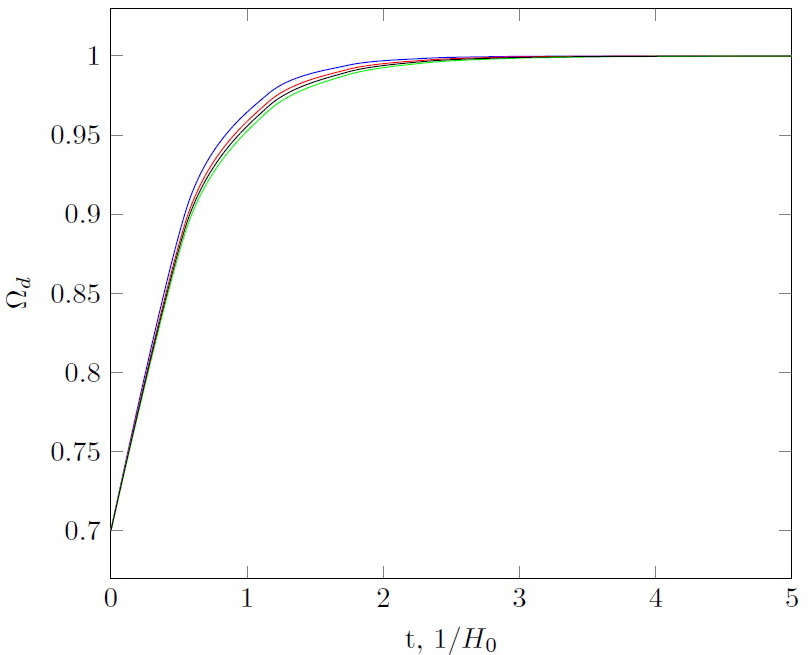}
    \caption{Same as in Figure~\ref{fig11}, but for $C=1.1$, $d=0$. For $\gamma<1$, the universe expands asymptotically according to the de Sitter law. $\gamma \geq 1$ corresponds to $H\rightarrow 0$.}
    \label{fig17}
\end{figure}

Finally, let us turn to the case $C>1$. Without interaction for $\gamma=1$, the state parameter asymptotically tends to a value greater than $-1$ (Figure~\ref{fig17}). At $\gamma>1$, the value of the state parameter reaches a minimum $w_{min}>-1$ and then begins to grow. At $\gamma<1$, there are two possibilities, namely the value of $w_{de}$ slowly ``skips down'', tending to $-1$ at $t\rightarrow\infty$ or after minimum $w_{de}\rightarrow -1$ below. The Hubble parameter decreases with time and there are no singularities in the future. Only for $\gamma<1$ do we have de Sitter asymptotical expansion.

\begin{figure}[h]
\centering 
    \includegraphics[scale=0.25]{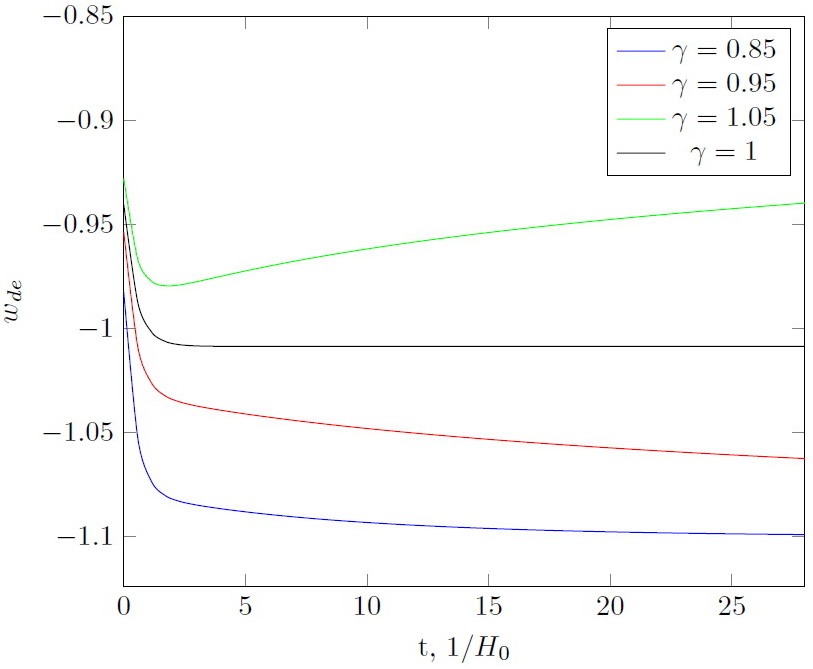}\includegraphics[scale=0.25]{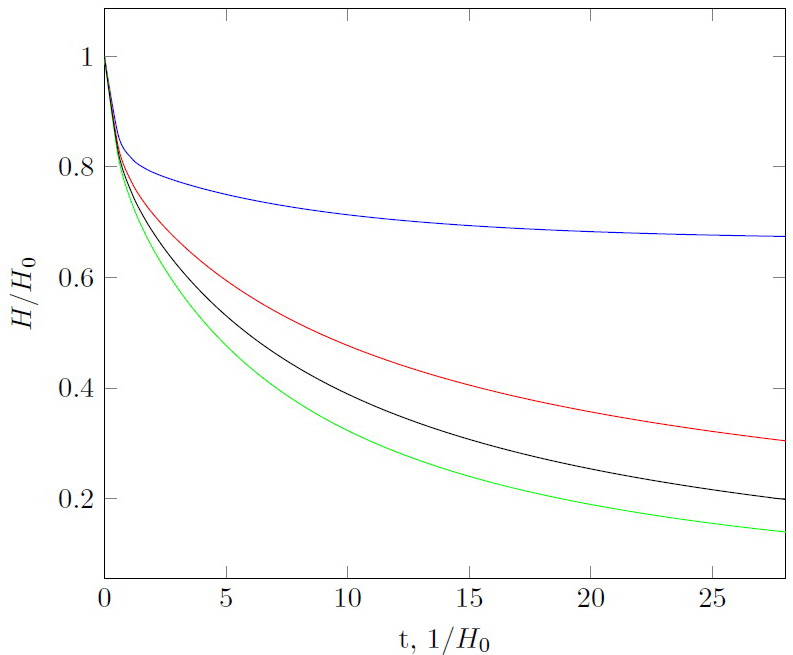}\\
    \includegraphics[scale=0.25]{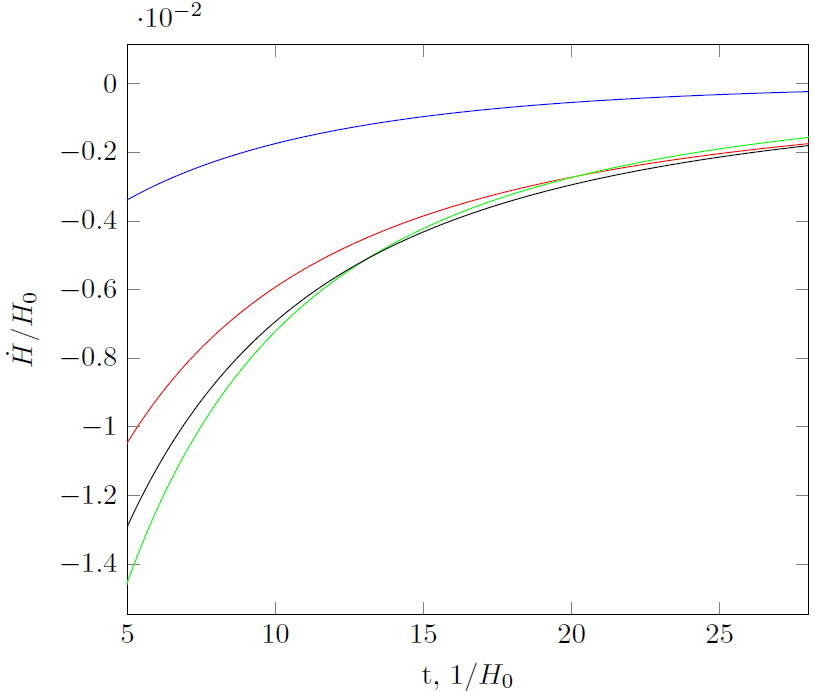}\includegraphics[scale=0.25]{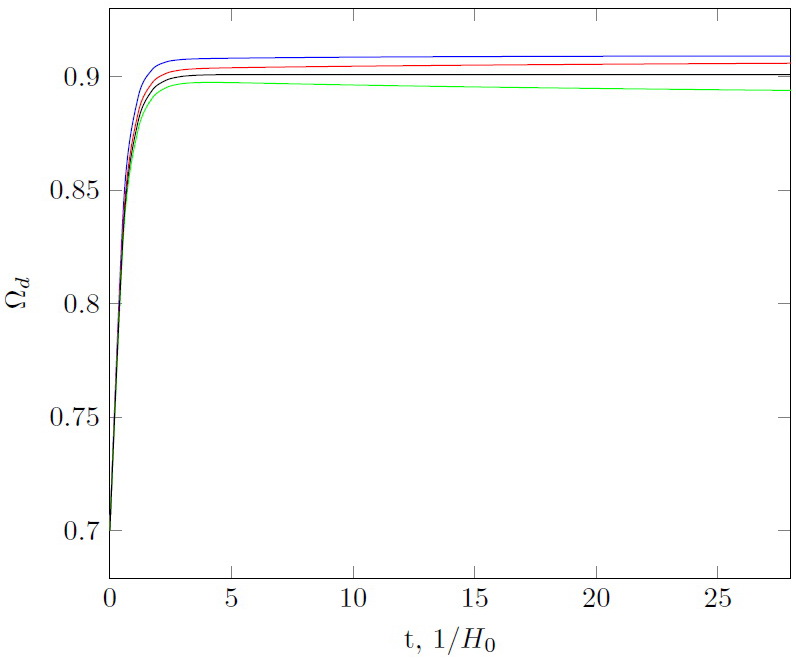}
    \caption{Same as in Figure~\ref{fig11}, but for $C=1.1$, $d^2=0.1$. Evolution of the universe is the same as without interaction but for quasi-de Sitter expansion $w_{de}\rightarrow w_0<-1$.}
    \label{fig18}
\end{figure}

Interaction at $\gamma=1$ leads to the fact that the asymptotic value of the parameter $w_{de}$ {over long periods} 
can be less than $-1$, but the Hubble parameter decreases, slowly tending to zero (see Figure~\ref{fig18}). For $\gamma<1$, we have de Sitter expansion with $w_{de}<-1$.  It is interesting to note that for $\gamma>1$ a double crossing of the phantom divide line can occur. The Hubble parameter decreases and $H\rightarrow 0$ for $t\rightarrow 0$.

\vspace{-12pt}

\section{Conclusions}

Let us briefly consider some another choices for the function $Q$.
\begin{equation}
Q=3d^2 H \rho_{de}^{\alpha}\rho^{\beta},\quad \rho=\rho_m+\rho_{de},
\end{equation}
$$
\alpha + \beta = 1.
$$

Our analysis shows that for various $\alpha$ and $\beta$, the evolution of the universe in principle does not contain principal peculiarities in comparison with the simple case $\alpha=1$, $\beta=0$ considered above.

For
 \begin{equation}
Q=3d^2 H (\rho_{de}-\rho_{m}),
\end{equation}
the situation is similar. The explanation is very simple. From previous calculations, one can see that the fraction of matter tends to a constant value and therefore $\rho_{de}-\rho_m\approx \delta \rho_{de}$, where $\delta<1$. This model is close to that considered above with the slightly redefined parameter~$d^2$.  

In the case of
 \begin{equation}
Q=3d^2 H \rho_{de}^{\alpha}\rho_{m}^{\beta},\quad \alpha + \beta = 1,
\end{equation}
the situation is more interesting. The big rip singularity for $C<1$ and $\gamma=1$ can be eliminated due to the interaction for $\alpha=2$, $\beta=-1$.

We investigated models of Tsallis holographic dark energy with inclusion of interaction between dark energy and matter in the form of a function $Q=3d^2H\rho_{de}$. Our calculations show that the interaction can lead to phantomization for dark energy but without a singularity in the future as one can expect for phantom energy with a constant value of $w_{de}=p_{de}/\rho_{de}$. The established equilibrium between holographic energy density and matter density cancels the singularity although the asymptotical value of $w_{de}$ can be less than $-1$. 
Another interesting issue is that for $C<1$, the future big rip singularity is eliminated due to the interaction. In this case, $w_{de}$ tends to a constant value $w_{0}<-1$ but the Hubble parameter slowly decreases. For $\gamma>1$, there is a critical value of the coupling parameter $d^2$ such that big rip singularity does not occur.
For $C>1$ and $\gamma\geq 1$, the``quintessence'' with $w<-1$ due to the interaction is possible: the Hubble parameter decreases and the universe expansion decelerates.

{We consider} 
 the epoch of late acceleration in our paper. Note that, in a similar fashion, following the approach developed in
\cite{Nojiri:2020wmh} one can consider unification of inflation with dark energy within Tsallis HDE.

{In conclusion,} 
 we need to say a few words about the coincidence problem in the considered model. In the $\Lambda$CDM model, the observed fact that the present fractions of dark energy and dark matter are of the same order of magnitude shows that we are currently living in a very special period. For Tsallis HDE without interaction, we also need to explain why in the present epoch dark energy and matter density are close to each other. However, interaction in some cases changes the behavior of the relation $\rho_{de}/\rho_m$ dramatically. For the case of $C=1$ and for appropriate $d^2$, we have a quasi-de Sitter evolution, but the fraction of dark energy tends to $\Omega_{de}<1$ (see Figure~\ref{fig12}). Moreover, this limit can be close to the current value and therefore there is nothing surprising in the fact $\rho_{de}/\rho_{m}\sim \mathcal{O}(1)$. Only in the past does $\rho_{m}>>\rho_{de}$. Even for a scenario with big rip singularity in the future, interaction can lead to that for most of cosmological evolution $\rho_{de}/\rho_{m}\sim \mathcal{O}(1)$ (Figure~\ref{fig15}, case $d^2=0.2$). The same picture is repeated for {quintessence-mimicking} 
 THDE with interaction (C > 1).         


\textbf{Acknowledgments}. {This research was funded by the Ministry of Education and Science (Russia), project 075-02-2021-1748.}

\end{document}